\documentclass[aps,showpacs,preprint,amsmath,amssymb]{revtex4}

\usepackage{epsfig}
\usepackage{graphicx}
\usepackage{bm}

\begin{document}

\title{Theoretical Construction of 1D anyon models}
\author{Ren-Gui Zhu}
\affiliation{Department of Modern Physics, University of Science and
Technology of China, Hefei, 230026, P.R.China}
\author{An-Min Wang}
\email{anmwang@ustc.edu.cn} \affiliation{Department of Modern
Physics, University of Science and Technology of China, Hefei,
230026, P.R.China}

\begin{abstract}
One-dimensional anyon models are renewedly constructed by using path
integral formalism. A statistical interaction term is introduced to
realize the anyonic exchange statistics. The quantum mechanics
formulation of statistical transmutation is presented.
\end{abstract}


\maketitle
The theory of anyons is used to describe the particles with
fractional exchange statistics\cite{Lein,Wilc,Hald}, which has
provided a successful explanation for the fractional quantum Hall
effect(FQHE)\cite{Halp}. Nowadays, it has been widely applied in
condensed matter physics and quantum computation. For two
dimensions(2D), the exchange statistics is well defined. But for one
dimension(1D), because two particles cannot be interchanged without
collision, the intrinsic statistics is inextricably mixed up with
the local interactions, this make the exchange statistics for one
dimension cannot be uniquely defined. However,
experiments\cite{1Dex} have demonstrated the possibility of
confining atoms to one dimension, so both conceptually and actually,
the 2D anyons can be confined to move in one dimension. Recently, a
number of researches on 1D anyons have been
reported\cite{Batch,Gira,Cala,Patu}.

In a seminal work, Kundu\cite{Kundu} defined a 1D anyon field
operator $\hat{\psi}_A(x)$ in terms of the Bose field operator
$\hat{\psi}_B(x)$ through a gauge transformation:
$\hat{\psi}_A(x)=e^{-i\kappa\int_{-\infty}^xdx'\hat{\rho}(x')}\hat{\psi}_B(x)$,
where
$\hat{\rho}(x)=\hat{\psi}_A^{\dagger}(x)\hat{\psi}_A(x)=\hat{\psi}_B^{\dagger}(x)\hat{\psi}_B(x)$
is the number density operator. The commutation relations are
$\hat{\psi}_A(x)\hat{\psi}_A^{\dagger}(x')=e^{-i\kappa\epsilon(x-x')}\hat{\psi}_A^{\dagger}(x')\hat{\psi}_A(x)+\delta(x-x')$,
and
$\hat{\psi}_A(x)\hat{\psi}_A(x')=e^{i\kappa\epsilon(x-x')}\hat{\psi}_A(x')\hat{\psi}_A(x)$,
where $\epsilon(x)=+1(-1)$ for $x>0(x<0)$, and $\epsilon(0)=0$. An
alternative definition proposed by Girardeau\cite{Gira} is in terms
of the Fermi field operator:
$\hat{\psi}_A(x)=e^{-i\kappa\int_{-\infty}^xdx'\hat{\rho}(x')}\hat{\psi}_F(x)$,
where
$\hat{\rho}(x)=\hat{\psi}_A^{\dagger}(x)\hat{\psi}_A(x)=\hat{\psi}_F^{\dagger}(x)\hat{\psi}_F(x)$.
The commutation relations become
$\hat{\psi}_A(x)\hat{\psi}_A^{\dagger}(x')+e^{-i\kappa\epsilon(x-x')}\hat{\psi}_A^{\dagger}(x')\hat{\psi}_A(x)=\delta(x-x')$,
and
$\hat{\psi}_A(x)\hat{\psi}_A(x')+e^{i\kappa\epsilon(x-x')}\hat{\psi}_A(x')\hat{\psi}_A(x)=0$.
Note that in this case, the exclusion principle
$\hat{\psi}_A^2(x)=[\hat{\psi}_A^{\dagger}(x)]^2=0$ is satisfied
automatically.

From the above commutation relations, the basic exchange symmetry of
the $N$-body anyonic wave function
$\Psi_A(x_1,...,x_N)=\langle0|\hat{\psi}_A(x_1)...\hat{\psi}_A(x_N)|N\rangle$
by transposing only the adjacent $x$'s can be
obtained\cite{Gira,Kundu}:
\begin{equation}\label{eq01}
\Psi_A(...,x_i,x_{i+1},...)=\pm
e^{-i\kappa\epsilon(x_i-x_{i+1})}\Psi_A(...,x_{i+1},x_i,...),
\end{equation}
where we have written the exchange symmetry of the two kinds of
anyon models in one equation, utilizing the sign $\pm$. The plus
sign is for the Kundu's anyons and the minus sign is for the
Girardeau's. For exchanging two arbitrary particles, the exchange
symmetry expression\cite{Batch,Gira,Kundu} is $\Psi_A(...,x_i,...,
x_j,...)= \pm e^{-i\theta}\Psi_A(...,x_j,..., x_i,...)$ with
$\theta=\kappa
[\sum_{k=i+1}^j\epsilon(x_i-x_k)-\sum_{k=i+1}^{j-1}\epsilon(x_j-x_k)]$,
which can be derived from the iteration of the basic exchange
symmetry Eq.(\ref{eq01}).

One sees that the phase factor appearing under the exchange of two
particles also depends on the coordinates of the particles between
them. This is a distinct feature of 1D anyons, because two particles
in a line can not pass each other without exchange. This fact
hindered the early attempts at direct introduction of the 1D anyons
as charge-flux composites\cite{Rabe,Agli}, or as a flux-carrying
boson (or fermion), which is known as statistical transmutation. In
two dimensions, this problem has been well solved, the flux is
called the Chern-Simons flux\cite{Khare}.

In this paper, we try to seek for a theoretical construction of 1D
anyon models in order to obtain a unitive description for 1D anyon
models, at least, for Kundu's and Girardeau's, and we have found an
analogical quantum mechanics formulation of the statistical
transmutation for 1D particles as in two dimensions, including an
auxiliary field which plays the role of the Chern-Simons field in 2D
case. The key step in our theoretical construction is to correctly
give a 1D statistical interaction term that can realize the anyonic
exchange statistics presented above.

\begin{figure}[h]
\begin{center}
\includegraphics[scale=0.5]{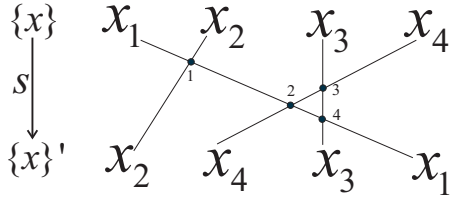}
\caption{\label{anyonicfactor}The crossing points $1$, $2$, $3$ and
$4$ give the factor $\pm e^{-i\kappa\epsilon(x_1-x_2)}$, $\pm
e^{-i\kappa\epsilon(x_1-x_4)}$, $\pm e^{-i\kappa\epsilon(x_3-x_4)}$
and $\pm e^{-i\kappa\epsilon(x_1-x_3)}$ respectively, so their
product is
$F_S(\{x\})=e^{-i\kappa[\sum_{i=2}^4\epsilon(x_1-x_i)+\epsilon(x_3-x_4)]}$.}
\end{center}
\end{figure}

To make our construction of the statistical interaction term look to
be natural, it is necessary to write the exchange symmetry of the
anyonic wave function in a compact form. The more general form of
exchange symmetry in principle can be obtained from the iteration of
the basic relation, but it is still not very convenient. Here we
present a most general exchange symmetry expression written in a
compact form:
\begin{equation}
\Psi_A(\{x\})=F_S(\{x\})\Psi_A(\{x\}'),
\end{equation}
where $\{x\}$ is a permutation of $x_1,...,x_N$, and $S$ is a
permutation operator that takes $\{x\}$ into $\{x\}'$,
i.e.,$\{x\}'=S\{x\}$. The phase factor $F_S(\{x\})$ can be obtained
from a rule : write down $\{x\}$ in a horizontal line, and under it
write down $\{x\}'$, use a straight line to connect both the $x_j$'s
($j=1,...,N$) in the two rows, then every crossing point of the two
lines corresponding to $x_i$ and $x_j$ contributes a phase factor
$\pm e^{-i\kappa\epsilon(x_i-x_j)}$, where $x_i$ stands on the left
side of $x_j$ in the $\{x\} $ row. Then $F_S(\{x\})$ is the product
of all these factors:
\begin{equation}
F_S(\{x\})=e^{-i\kappa\sum_{(i,j),S}\epsilon(x_i-x_j)}
\end{equation}
for Kundu's anyons, and
\begin{equation}\label{eq07}
F_S(\{x\})=(-1)^ne^{-i\kappa\sum_{(i,j),S}\epsilon(x_i-x_j)}
\end{equation}
for Girardeau's anyons. Here, the summation $\sum_{(i,j),S}$ means
only counting the crossing points produced by permutation $S$, and
$n$ is the number of the crossing points. We mention here that the
rule above is very similar to the rule put forward by Lieb and
Liniger\cite{Lieb} for writing the scattering amplitude in their
Bethe ansatz solution of 1D $\delta$-Bose gas. In
Fig.\ref{anyonicfactor}, an example of writing the phase factor is
given. One can check that using this rule, the above phase factor
$\pm e^{-i\theta}$ can be conveniently obtained. A reasonable
explanation for this rule will be given in the following.

From the compact expression of the exchange symmetry, one can see
that the exchange phase factor is only determined by the permutation
of the particles' coordinates. In fact, there are many different
paths in the configuration space leading to the transposition from
$\{x\}$ to $\{x\}'=S\{x\}$, but all the different paths leading to
$\{x\}'$ will give the same exchange phase factor. There are two
reasons for this fact: $(a)$ According to the basic exchange
symmetry, if two particles are exchanged for even times, they will
contribute nothing to the exchange phase factor. If they are
exchanged for odd times, the result is the same as their being
exchanged only once. $(b)$ During the exchange process, changing the
order of the adjacent exchanges will not affect the final phase
factor. Based on $(a)$ and $(b)$, we can only consider the direct
path comprised by a series of adjacent exchanges, in which two
particles exchange at most once. All the other paths leading to
$\{x\}'$ are equivalent to this direct path in view of producing the
same exchange phase factor. So to obtain the phase factor, what all
we need to know is how many and what adjacent exchanges happened
during the permutation, which give a reasonable explanation of the
above rule for writing the phase factor.

From the above analysis, one see that all the paths from $\{x\}$ to
$\{x\}'$ can be thought of as belonging to an equivalent class,
which can be denoted by $S$ and give the same exchange phase factor.
This implies that the anyonic statistics here may be possible to be
elucidated by Feynman's path integral formalism\cite{Feyn} just as
done in Ref.\cite{Laidlaw,Wu} for two-dimensional fractional
statistics. Let's consider a one-dimensional $N$-body system with
Lagrangian $L=L_0+L_s$. Here $L_s$ is constructed as
\begin{equation}
L_s=\frac{\hbar\kappa}{2}\frac{d}{dt}\left[\sum_{i<j}\epsilon(x_i-x_j)+l(x_1,...,x_N)\right],
\end{equation}
where the $l(x_1,...,x_N)$ is a function whose contribution to the
path integral is only determined by the permutation of the
coordinates, as well as the function $\sum_{i<j}\epsilon(x_i-x_j)$.
We will show below that this form of $L_s$ is the statistical
interaction term for 1D anyons. For Kundu's anyons, $l(x_1,...,x_N)$
can be any constant or zero. And for Girardeau's anyons, we make
$l(x_1,...,x_N)=\frac{\pi}{\kappa}A(x_1,...,x_N)$, where
$A(x_1,...,x_N)=\prod_{i<j}$sgn$(x_i-x_j)$ is just the mapping
function introduced by Girardeau for the theory of Fermi-Bose
mapping\cite{Gira2}, and takes the value $\pm1$ according to the
parity of the permutation of the coordinates. Theoretically, other
forms of $l(x_1,...,x_N)$ is also allowed, which means that there
maybe exist some other kinds of anyons besides Kundu's and
Girardeau's.

According to Feynman's path integral formalism\cite{Feyn}, the
transition amplitude from $(\textbf{x},t)$ to $(\textbf{x}',t')$ in
the configuration space is given by\cite{Khare}
\begin{equation}\label{eq02}
\langle\textbf{x}',t'|\textbf{x},t\rangle\propto\sum_{all\
paths}\exp\left(\frac{i}{\hbar}\int_{t}^{t'}Ldt\right),
\end{equation}
where $\textbf{x}$ is an initial configuration point with fixed
permutation of the particles, and $\textbf{x}'$ is the same
configuration point, however, with an uncertain permutation of the
particles, because of the indistinguishability of the particles
during the evolvement, which means that $\textbf{x}'$ can be
$S\textbf{x}$ for any element $S$ in the permutation group $ S_N$.
Without any loss of generality, we can make
$\textbf{x}=(x_1,x_2,...,x_N)$.

As we have stated above, the contribution of $L_s$ to the the path
integral, i.e., the right side of Eq.(\ref{eq02}) is only determined
by the final permutation of the coordinates of the particles, so
Eq.(\ref{eq02}) can be rewritten as a weighted sum over all the
permutations, following the formulation given in\cite{Wu}:
\begin{equation}\label{eq03}
\langle\textbf{x}',t'|\textbf{x},t\rangle=\sum_{S\in
S_N}\chi(S,\textbf{x})\sum_{path\in
S}\exp\left(\frac{i}{\hbar}\int_t^{t'}Ldt\right),
\end{equation}
where $\sum_{path\in S}$ means the summation takes over all the
paths leading to the same permutation $S\textbf{x}$, and the weight
factor is
\begin{eqnarray}\label{eq04}
\chi(S,\textbf{x})&=&\exp\left(\frac{i}{\hbar}\int_t^{t'}L_sdt\right)\nonumber\\
&=&\exp\left[i\frac{\kappa}{2}\sum_{i<j}\epsilon(x_i-x_j)|_{\textbf{x}}^{S\textbf{x}}
+\frac{i\kappa}{2}l(x_1,...,x_N)|_{\textbf{x}}^{S\textbf{x}}\right].
\end{eqnarray}
From Eq.(\ref{eq03}), it looks as if that the permutation group
$S_N$ is isomorphic to the first homotopy group of the configuration
space of this 1D $N$-body system. But it is not the fact, because in
the concept of homotopy group, the paths are classified into
homotopy classes by whether they can be deformed into each other,
but here all the paths are classified by whether they can lead to
the same permutation. However, it is clear that the permutation
group $S_N$ takes the equivalent role in this one-dimensional case
as the homotopy group in two-dimensional case\cite{Wu}, of course
under the condition of a given permutation $\textbf{x}$ of the
particles' coordinates as the initial configuration. So we hope that
$\chi(S,\textbf{x})$ be a phase factor and equivalent to
$F_S(\{x\})$, provided $\textbf{x}=\{x\}$ as an initial permutation,
i.e.:
\begin{equation}\label{eq05}
\chi(S,\textbf{x})=F_S(\{x\}).
\end{equation}
It is not difficult to check the correctness of this equation. In
fact, this is the original idea of our construction of the
statistical term $L_s$. Here, we give a brief explanation of
Eq.(\ref{eq05}) in the following. For Kundu's anyons, we let
$l(x_1,...,x_N)$ be any constant or zero. Inspecting the
Eq.(\ref{eq04}) tell us that if two particles with initial
coordinates, e.g. $x_i,x_j$ with $i<j$, are not exchanged during the
process to the new permutation $S\textbf{x}$, they will contribute
nothing to $\chi(S,\textbf{x})$. But if they are exchanged, their
contribution will be $e^{-i\kappa\epsilon(x_i-x_j)}$. This coincides
with our rule for writing the phase factor $F_S(\{x\})$. While for
Girardeau's anyons, we take
$l(x_1,...,x_N)=\frac{\pi}{\kappa}A(x_1,...,x_N)$, which is
responsible for the factor $(-1)^n$ in the right side of
Eq.(\ref{eq07}), because the number of crossing points $n$ can be
replaced by the parity of the permutation.

Subsequently, let's consider the quantum mechanics of 1D anyons
following the formalism in Ref.\cite{Ezaw} for 2D anyons. For
simplicity, we only present the formulation for Kundu's anyons in
the following. The formulation for other anyons can be obtained in
the same way by considering the additional term $l(x_1,...,x_N)$.
The Lagrangian for ordinary bosons can be written as:
\begin{equation}
L_0=\frac{m}{2}\sum_{i=1}^N\dot{x_i}^2-V(x_1,...,x_N).
\end{equation}
The properties of Kundu's anyons can be obtained if the statistical
term $L_s$ is included:
\begin{equation}
L=L_0+\hbar\kappa\sum_{i<j}\delta(x_i-x_j)(\dot{x_i}-\dot{x_j}),
\end{equation}
where the use of identity
$\epsilon(x_i-x_j)=\theta(x_i-x_j)-\theta(x_j-x_i)$ has been made.
$\theta(x)$ is the Heaviside's step function. The conjugate momentum
is
\begin{equation}
p_k=m\dot{x_k}+eA_k,
\end{equation}
where
\begin{equation}
eA_k=\hbar\kappa\left[\sum_{j=k+1}^{N}\delta(x_k-x_j)-\sum_{j=1}^{k-1}\delta(x_j-x_k)\right]
\end{equation}
is introduced for brevity. But soon, we will see that $eA_k$ has
further significance: the parameter $e$ can be regarded as a charge,
and $A_k$ can be regarded as an auxiliary field potential like the
Chern-Simons field potential in two dimensions (see the Hamiltonian
below). Using the Legendre transformation $H=\sum_kp_k\dot{x_k}-L$,
and after a straightforward derivation, one can finally obtain the
Hamiltonian in a compact form:
\begin{equation}
H=\frac{1}{2m}\sum_k(p_k-eA_k)^2+V(x_1,...,x_N),
\end{equation}
The first quantization form of which can be obtained just by taking
$p_k\to-i\hbar\partial_{x_k}$. It is the Hamiltonian of a
flux-carrying bosons system with complicated interaction, however,
it also can be regarded as a free anyon system apart from the
interaction $V(x_1,...,x_n)$. This can be quickly seen if we operate
a phase transformation on the Schr\"{o}dinger equation
\begin{equation}
[-\frac{\hbar^2}{2m}\sum_k(\partial_{x_k}-i\frac{e}{\hbar}A_k)^2+V(x_1,...,x_N)]\Psi(x_1,...,x_N)=E\Psi(x_1,...,x_N).
\end{equation}
The phase transformation is
\begin{equation}\label{eq06}
\tilde{\Psi}=e^{ief(x_1,...,x_N)/\hbar}\Psi,
\end{equation}
where $f(x_1,...,x_N)$ is requested to satisfy
$\partial_{x_k}f=-A_k$, and can be chosen as
\begin{equation}
f(x_1,...,x_N)=-\frac{\hbar\kappa}{2e}\sum_{i<j}\epsilon(x_i-x_j).
\end{equation}
Consequently the transformed Hamiltonian is
\begin{equation}
\tilde{H}=-\frac{\hbar^2}{2m}\sum_k\partial_{x_k}^2+V(x_1,...,x_N),
\end{equation}
which represents a pure 1D anyons system. The corresponding wave
function $\tilde{\Psi}$ has anyonic exchange symmetry. Because
$\Psi$ is a Bose wave function, the phase transformation
Eq.(\ref{eq06}) is just the anyon-boson mapping presented in Kundu's
paper\cite{Kundu}. We note here that in an earlier work\cite{Rabe},
the attempt to establish this kind of statistical transmutation
failed, because the Hamiltonian there didn't include the potential
like $A_k$ here.

For Girardeau's anyons, because of the intrinsic property of
exclusion principle, the interaction term including $\delta$
function must vanish, which make the formation of Hamiltonian be of
no difference from the Bose or Fermi one\cite{Gira}. So the
transformation of the Hamiltonian is a trivial problem, knowing the
formulation of anyon-fermion(or boson) mapping is enough for
statistical transmutation. If $\Psi$ is a Fermi (or Bose) wave
function, the phase transformation Eq.(\ref{eq06}) is just the
anyon-fermion(or boson) mapping for Girardeau's anyons\cite{Gira}.

In conclusion, we have obtained a theoretical construction for 1D
anyon models based on a form of statistical interaction term.
According to the path integral formalism, this interaction term can
successfully recover the anyonic exchange symmetry of the known
anyons defined by Kundu and Girardeau. Furthermore, it theoretically
provide a unitive description for Kundu's and Girardeau's, and
perhaps for more extensive 1D anyon models beyond them. The
statistical transmutation in quantum mechanics formalism are also
presented, the known anyon-fermion (or boson) mapping relations are
successfully recovered.

We are grateful to all the collaborators in our quantum theory group
at the Institute for Theoretical Physics of our university. This
work was supported by the National Nature Science Foundation of
China under Grant No. 60573008.


\begin{thebibliography}{99}
\bibitem{Lein}J. M. Leinaas and J. Myrherim, Nuovo Cimento B
\textbf{37}, 1 (1977).
\bibitem{Wilc}F. Wilczek, Phys. Rev. Lett. \textbf{48}, 1144 (1982); \textbf{49}, 957 (1982).
\bibitem{Hald}F. D. M. Haldane, Phys. Rev. Lett. \textbf{67}, 937
(1991).
\bibitem{Halp}B. I. Halperin, Phys. Rev. Lett. \textbf{52}, 1583
(1984); \emph{Fractional Statistics and Anyon Superconductivity},
edited by F. Wilczek (World Scientific 1990).
\bibitem{1Dex}A. G$\ddot{o}$rlitz et al., Phys. Rev. Lett.
\textbf{87}, 130402 (2001); F. Schreck et al., Phys. Rev. Lett.
\textbf{87}, 080403 (2001); N. Wada et al., Phys. Rev. Lett.
\textbf{86}, 4322 (2001); W. H$\ddot{a}$nsel et al., Nature
\textbf{413}, 498 (2001); M. Greiner et al., Phys. Rev. Lett.
\textbf{87}, 160405 (2001).
\bibitem{Batch}M. T. Batchelor, X.-W. Guan and N. Oelkers, Phys.
Rev. Lett. \textbf{96}, 210402 (2006); M. T. Batchelor and X.-W.
Guan, Phys. Rev. B \textbf{74}, 195121 (2006); M. T. Batchelor,
X.-W. Guan and J.-S. He, J. Stat. Mech. P03007 (2007).
\bibitem{Gira}M. D. Girardeau, Phys. Rev. Lett. \textbf{97}, 100402
(2006).
\bibitem{Cala}P. Calabrese and M. Mintchev, Phys. Rev. B
\textbf{75}, 233104 (2007).
\bibitem{Patu}O. I. P$\hat{a}$tu, V. E. Korepin and D. V. Averin,
arXiv:0707.4520.
\bibitem{Kundu}A. Kundu, Phys. Rev. Lett. \textbf{83}, 1275 (1999).
\bibitem{Rabe}S. J. B. Rabello, Phys. Lett. B \textbf{363}, 180
(1995); Phys. Rev. Lett. \textbf{76}, 4007(1996); Phys. Rev. Lett.
\textbf{77}, 4851 (1996).
\bibitem{Agli}U. Aglietti, L. Griguolo, R. Jackiw, S. Y. Pi and D.
Semimara, Phys. Rev. Lett. \textbf{77}, 4406 (1996).
\bibitem{Khare}A. Khare, \emph{Fractional Statistics and Quantum
Theory} (World Scientific 2005), and reference therein.
\bibitem{Lieb}E. H. Lieb and W. Liniger, Phys. Rev. \textbf{130},
1605 (1963).
\bibitem{Feyn}R. P. Feynman and A. R. Hibbs, Quantum Mechanics and
Path Integrals(McGraw-Hill, New York, 1965).
\bibitem{Laidlaw}M. G. G. Laidlaw and C. M. De Witt, Phys. Rev. D
\textbf{3}, 1375 (1971).
\bibitem{Wu}Yong-Shi Wu, Phys. Rev. Lett. \textbf{52}, 2103 (1984).
\bibitem{Gira2}M. Girardeau, J. Math. Phys. \textbf{1}, 516 (1960).
\bibitem{Ezaw}Z. F. Ezawa, \emph{Quantum Hall Effects: Field Theoretical Approach
and Related Topics}(World Scitific 2000).
\end{thebibliography}
\end{document}